\documentclass{elsart}
\usepackage[dvips]{graphicx}
\usepackage{amssymb}

\begin{document}

\renewcommand{\baselinestretch}{1}
\normalsize
\begin{center}
{\large \bf Measurement of radon concentrations at Super-Kamiokande}

\end{center}

\small

\begin{center}
\newcounter{foots}

%FIRST AUTHORs (radon group)
Y.Takeuchi$^a$, K.Okumura$^a$, T.Kajita$^a$, S.Tasaka$^g$,
M.Nemoto$^t$, Y.Fukuda$^a$, 
\addtocounter{foots}{1}
H.Okazawa$^{o,\fnsymbol{foots}}$,
%
%ICRR
T.Hayakawa$^a$,
K.Ishihara$^a$, H.Ishino$^a$, Y.Itow$^a$,
J.Kameda$^a$, S.Kasuga$^a$, K.Kobayashi$^a$, Y.Kobayashi$^a$, 
Y.Koshio$^a$,   
M.Miura$^a$, M.Nakahata$^a$, S.Nakayama$^a$, Y.Obayashi$^a$,
A.Okada$^a$, N.Sakurai$^a$,
M.Shiozawa$^a$, Y.Suzuki$^a$, H.Takeuchi$^a$,
Y.Totsuka$^a$, S.Yamada$^a$,
%
%Boston U
M.Earl$^b$, A.Habig$^b$, E.Kearns$^b$, 
M.D.Messier$^b$, K.Scholberg$^b$, J.L.Stone$^b$,
L.R.Sulak$^b$, C.W.Walter$^b$, 
%
%BNL
M.Goldhaber$^c$,
%Irvine
T.Barszczak$^d$, D.Casper$^d$, W.Gajewski$^d$,
W.R.Kropp$^d$,  S.Mine$^d$,
%W.R.Kropp$^d$,
L.R.Price$^d$, M.Smy$^d$, H.W.Sobel$^d$, 
M.R.Vagins$^d$,
%
%CSU
K.S.Ganezer$^e$, W.E.Keig$^e$,
%
%George Mason U
R.W.Ellsworth$^f$,
%
%Gifu U
%
%Hawaii U 
A.Kibayashi$^h$, J.G.Learned$^h$, S.Matsuno$^h$,
V.J.Stenger$^h$, D.Takemori$^h$,
%
%KEK
T.Ishii$^i$, J.Kanzaki$^i$, T.Kobayashi$^i$,
K.Nakamura$^i$, K.Nishikawa$^i$,
Y.Oyama$^i$, A.Sakai$^i$, M.Sakuda$^i$, O.Sasaki$^i$,
%
%Kobe U
S.Echigo$^j$, M.Kohama$^j$, A.T.Suzuki$^j$,
%
%Los Alamos
T.J.Haines$^{k,d}$,
%
%LSU
E.Blaufuss$^l$, B.K.Kim$^l$, R.Sanford$^l$, R.Svoboda$^l$,
%
%Maryland U
M.L.Chen$^m$,J.A.Goodman$^m$, G.W.Sullivan$^m$,
%
%
%SUNY
J.Hill$^n$, C.K.Jung$^n$, K.Martens$^n$, C.Mauger$^n$, C.McGrew$^n$,
E.Sharkey$^n$, B.Viren$^n$, C.Yanagisawa$^n$,
%
%Niigata U
W.Doki$^o$, M.Kirisawa$^o$, S.Inaba$^o$,
K.Miyano$^o$,
C.Saji$^o$, M.Takahashi$^o$, M.Takahata$^o$,
%
%Osaka U.
K.Higuchi$^p$, Y.Nagashima$^p$, M.Takita$^p$,
T.Yamaguchi$^p$, M.Yoshida$^p$, 
%
%Seoul National University
S.B.Kim$^q$, 
%Tohoku U.
M.Etoh$^r$, A.Hasegawa$^r$, T.Hasegawa$^r$, S.Hatakeyama$^r$,
K.Inoue$^r$, T.Iwamoto$^r$, M.Koga$^r$, T.Maruyama$^r$, H.Ogawa$^r$,
J.Shirai$^r$, A.Suzuki$^r$, F.Tsushima$^r$,
%
%Tokyo U
M.Koshiba$^s$,
%
%Tokai U
Y.Hatakeyama$^t$, M.Koike$^t$, K.Nishijima$^t$,
%
%TIT
H.Fujiyasu$^u$, T.Futagami$^u$, Y.Hayato$^u$, 
Y.Kanaya$^u$, K.Kaneyuki$^u$, Y.Watanabe$^u$,
%
%Warsaw U
D.Kielczewska$^{v,d}$, 
%
%U Washington
\addtocounter{foots}{1}
\addtocounter{foots}{1}
\addtocounter{foots}{1}
J.S.George$^{w,\fnsymbol{foots}}$, A.L.Stachyra$^w$,
\addtocounter{foots}{1}
L.L.Wai$^{w,\fnsymbol{foots}}$, 
R.J.Wilkes$^w$, K.K.Young$^{w,\dagger}$

\footnotesize \it

$^a$Institute for Cosmic Ray Research, University of Tokyo, Tanashi,
Tokyo 188-8502, Japan\\
$^b$Department of Physics, Boston University, Boston, MA 02215, USA  \\
$^c$Physics Department, Brookhaven National Laboratory, Upton, NY 11973, USA \\
$^d$Department of Physics and Astronomy, University of California, Irvine,
Irvine, CA 92697-4575, USA \\
$^e$Department of Physics, California State University, 
Dominguez Hills, Carson, CA 90747, USA\\
$^f$Department of Physics, George Mason University, Fairfax, VA 22030, USA \\
$^g$Department of Physics, Gifu University, Gifu, Gifu 501-1193, Japan\\
$^h$Department of Physics and Astronomy, University of Hawaii, 
Honolulu, HI 96822, USA\\
$^i$Institute of Particle and Nuclear Studies, High Energy Accelerator
Research Organization (KEK), Tsukuba, Ibaraki 305-0801, Japan \\
$^j$Department of Physics, Kobe University, Kobe, Hyogo 657-8501, Japan\\
$^k$Physics Division, P-23, Los Alamos National Laboratory, 
Los Alamos, NM 87544, USA. \\
$^l$Department of Physics and Astronomy, Louisiana State University, 
Baton Rouge, LA 70803, USA \\
$^m$Department of Physics, University of Maryland, 
College Park, MD 20742, USA \\
$^n$Department of Physics and Astronomy, State University of New York, 
Stony Brook, NY 11794-3800, USA\\
$^o$Department of Physics, Niigata University, 
Niigata, Niigata 950-2181, Japan \\
$^p$Department of Physics, Osaka University, Toyonaka, Osaka 560-0043, Japan\\
$^q$Department of Physics, Seoul National University, Seoul 151-742, Korea\\
$^r$Department of Physics, Tohoku University, Sendai, Miyagi 980-8578, Japan\\
$^s$The University of Tokyo, Tokyo 113-0033, Japan \\
$^t$Department of Physics, Tokai University, Hiratsuka, Kanagawa 259-1292, 
Japan\\
$^u$Department of Physics, Tokyo Institute of Technology, Meguro, 
Tokyo 152-8551, Japan \\
$^v$Institute of Experimental Physics, Warsaw University, 00-681 Warsaw,
Poland \\
$^w$Department of Physics, University of Washington,    
Seattle, WA 98195-1560, USA    \\
\end{center}

\normalsize
%%%%%% for submit %%%%%%%%%%%%%%%%%%%%%%%%%
%\renewcommand{\baselinestretch}{2}
%%%%%%%%%%%%%%%%%%%%%%%%%%%%%%%%%%%%%%%%%%%

\newpage

\begin{quote}
\begin{center}
\bf Abstract
\end{center}
Radioactivity from radon is a major background for observing 
solar neutrinos at Super-Kamiokande. 
In this paper, we describe the measurement of radon concentrations
at Super-Kamiokande, the method of radon reduction, and the radon 
monitoring system.
The measurement shows that the current low-energy event rate between 
5.0 MeV and 6.5 MeV
implies a radon concentration in the Super-Kamiokande water
of less than 1.4 mBq/m$^3$.

\end{quote}

\vspace{3mm}
\normalsize

\section{Introduction}

An important feature of the Super-Kamiokande (SK)
detector is its ability to measure the energy spectrum of the recoil
electrons from $^8$B solar neutrinos\cite{skflux,skspec}.
In order to study the complete energy spectrum,
it is desirable to lower the analysis energy threshold as far as possible.
The current trigger threshold at 50\% efficiency
is 4.6 MeV in the total energy of electrons,
and the current analysis threshold is 5.5 MeV.
These thresholds are essentially limited by the high rate of the 
background events
due to the beta decays of radon daughters remaining in the water of SK
and gamma-rays from the PMT glass and plastic sheets.

Radon concentration in the Kamiokande-III detector was about 0.5 Bq/m$^3$,
and its energy threshold was 7.0 MeV \cite{kam_solar,suichu_radon,icrr_an}.
For the SK detector, a more efficient radon-reduction system was designed,
and a high sensitivity radon sensor was developed
to monitor the radon concentration in the water of SK\cite{ri,nim_radon}.
We have also studied the response of the SK detector to the radon events
by using a purified water with a known amount of radon.
In this paper, we report the study of radon radioactivity at Super-Kamiokande.

\section{Water and air purification system}

\begin{figure}[b]
\center
\includegraphics[width=14cm,clip]{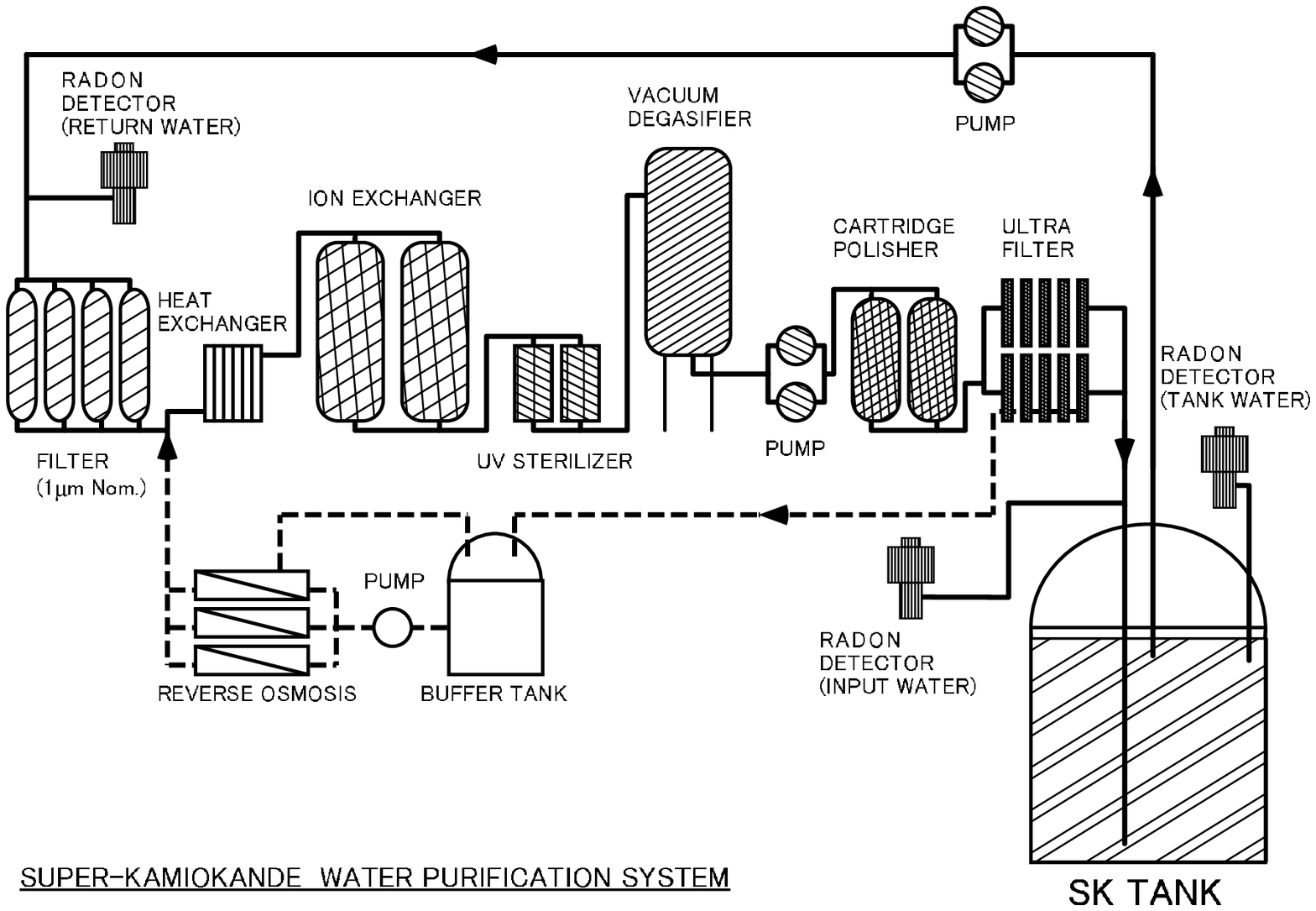}
\caption{
}
\label{water_system}
\end{figure}

The total 50,000 tons of purified water in the SK is 
circulated through a water purification system.
Figure~\ref{water_system} shows a schematic view of the water 
purification system.
The water flow rate is about 55 ton/hour in a closed cycle system.

Nominal 1 $\mu$m mesh filters and hollow fiber membrane filters are used 
to remove dust and relatively large particles in the water. 
These particles are though to be an important source of radon, because
they may contain radium, a parent atom of radon.
After the purification, a typical number of particles in the water is 
about 6 particles/m$\ell$ with a diameter larger than 0.2 $\mu$m.

An ion-exchanger and a cartridge polisher eliminate radium ions
in the same way as other heavy ions in the water.
After the cartridge polisher module, the resistivity of the water reaches 
almost the chemical limit of water, 18.24 M$\Omega \cdot $cm.

In order to remove radon gas dissolved in the water, 
a vacuum degasifier device is used.
The removal efficiency of radon gas in the water was measured
to be about 96\% at the Kamiokande site.

Air is also purified by an air purification system.
The purified air contains less radon, and is called ``Rn-reduced air''.
The Rn-reduced air is supplied to the space between the water surface and
the roof of the SK tank.
A total of 8 m$^3$ of activated carbon are used for removing radon 
in the air. 
Typical flow rate and dew point are 15 m$^3$/hour and --50 $^{\rm o}$C,
respectively.

The details about the water and air purification system will be
reported in Ref.~\cite{detector}.

\section{Radon monitoring system}

In order to monitor radon in the purified air and water, high sensitivity 
radon sensors were developed.
Radon is detected by electrostatic collection of the daughter nuclei 
of $^{222}$Rn and the energy measurement of the alpha decay 
with a PIN photodiode.
The details about the high sensitivity radon sensors are
reported in Ref.~\cite{nim_radon}.

\begin{figure}[b]
\center
\includegraphics[width=12cm,clip]{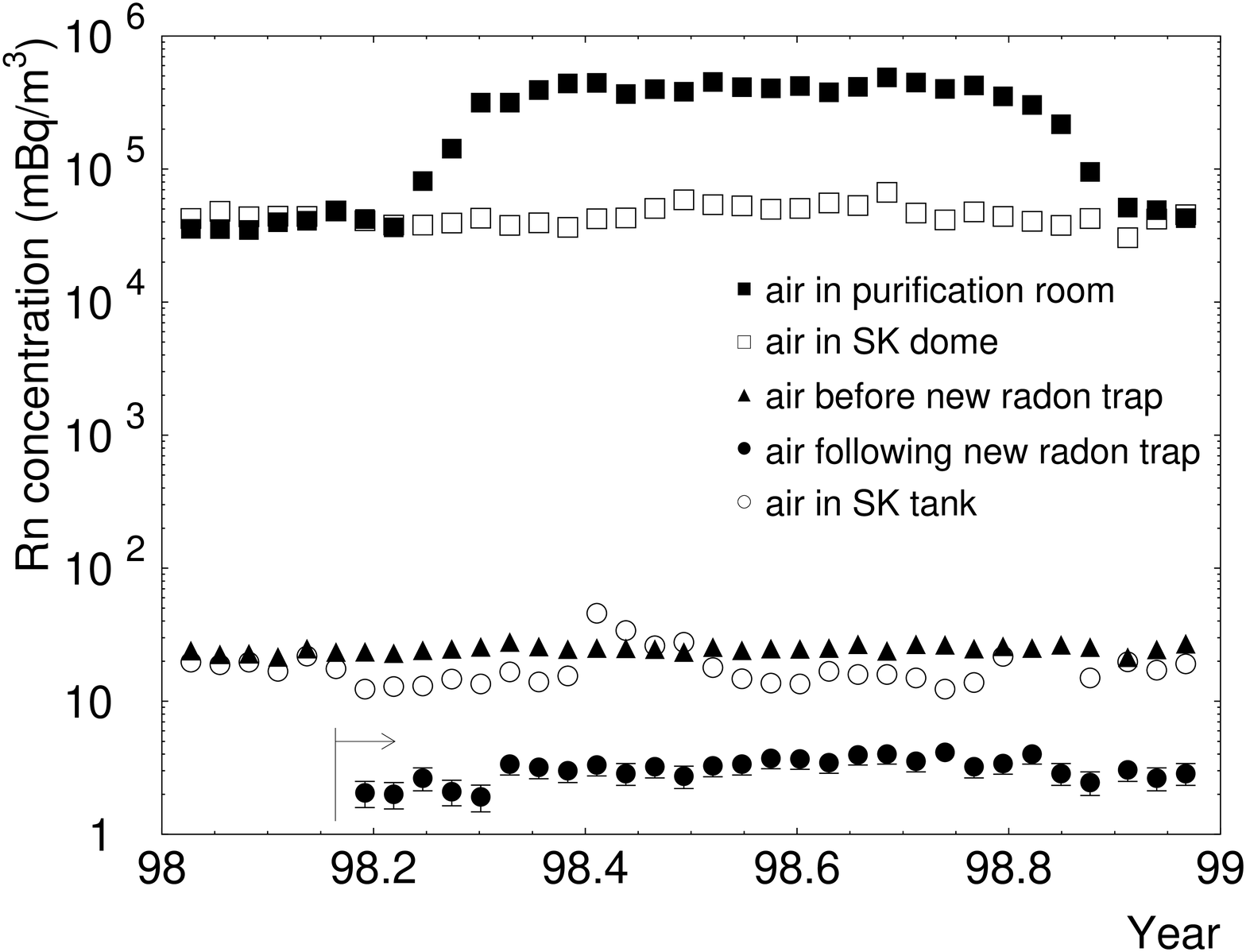}
\caption{
}
\label{rn_air}
\end{figure}

For monitoring radon in the air, five sensors are used.
They are connected to a workstation via a network, and radon 
concentrations are monitored in real-time.
Figure~\ref{rn_air} shows time variation of radon concentrations in the air
in 1998.

The ``air in purification room'' shown in Fig.~\ref{rn_air} shows
air in a room supplied with fresh air from outside the mine.
Although fresh air from outside is supplied to this room, there are
simple doors between this room and the mine tunnel, so that some
radon-rich mine tunnel air can leak in.
Typical radon concentrations in the purification room air in summer 
and in winter are 500 Bq/m$^3$ and 40 Bq/m$^3$, respectively.
A typical radon concentration in the mine tunnel air in summer is
about 1200 Bq/m$^3$.
The outside air blows directly into the mine tunnel near the water system
in the winter season, but the wind direction is reversed in the summer season.
Therefore, a variation of radon concentration in the purification room air
occurs seasonally.

The dome above the SK tank, called ``SK dome'', is coated with 
radon-tight plastic sheets
to prevent radon in the rock from entering into the air.
Fresh air from outside the mine is pumped into the SK dome at the rate of
$5 \sim 12$ m$^3$/hour, and there are air-tight doors between the SK dome
and the mine tunnel. 
Therefore, the radon concentrations in the SK dome are similar to
those of the outside air. 
A typical radon concentration in the dome air is 40 Bq/m$^3$.

A typical radon concentration after the air purification system (Rn-reduced air)
was about 20 mBq/m$^3$ before March, 1998 (``air before new radon trap'' 
in fig.~\ref{rn_air}).
In March 1998, an external radon trap was introduced into the air 
purification system. It consists of 50 $\ell$ of activated charcoal and a cooler.
The activated charcoal is cooled to --40 $^{\rm o}$C.
The radon concentration in the Rn-reduced air became $2 \sim 3$ mBq/m$^3$
after March 1998 (``air following new radon trap'' in fig.~\ref{rn_air}).

The radon concentration in the air occupied above the water surface 
in the SK tank, called ``SK-tank air'', was about 20 mBq/m$^3$ 
until March 1998, and about 11 mBq/m$^3$ after March 1998.
In the space between the water surface and the roof of the SK tank,
there are some materials emanating radon;
cables, dust, plastic sheets, and so on.
Therefore, the radon concentration increases for air going through this gap.
The bump of the radon concentration in SK-tank air in June was caused by
a malfunction of the air purification system.
The gap in the data around November was due to a malfunction
of real-time radon monitoring system. 

For monitoring radon concentrations in the purified water, 
a total of six high sensitivity radon sensors are used.
They are installed at three different locations,
two sensors for each location, to measure the radon concentrations of
return water from the SK tank, input water to the SK tank, 
and water in the SK tank. 
These locations are shown in Fig.~\ref{water_system}.

The return water is sampled for the measurement just before 
the water purification system.
The SK tank water is sampled 4 m  below the top of the SK tank.
The input water is sampled just before the SK tank
(just after the purification system).

\begin{figure}[b]
\center
\includegraphics[width=12cm,clip]{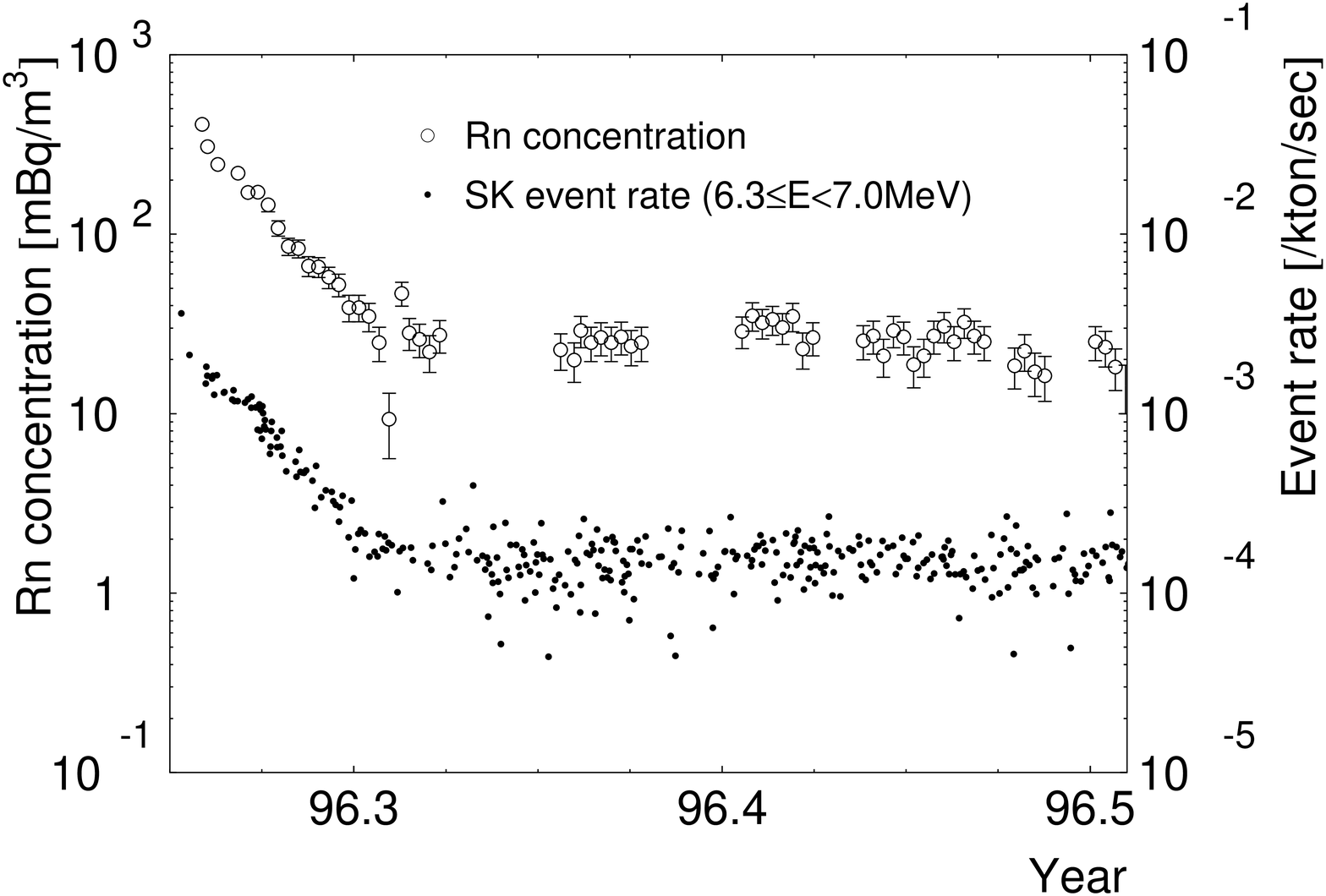}
\caption{
}
\label{rn_water}
\end{figure}

Figure~\ref{rn_water} shows the radon concentration in the return 
water and SK low-energy event rate at the beginning of the experiment.
The radon concentration at the beginning of the experiment
in April 1996 was essentially the same as Kamiokande-III, 500 mBq/m$^3$.
At that time, the low-energy event rate was consistent with being
caused by the high radon level alone. Following the sealing of the
SK detector and the beginning of purification circulation in March 1996,
the concentration dropped to less than 20 mBq/m$^3$.
There was another decrease in radon in July 1996, then 
the concentration became its current nominal value.

The current radon sensors are sensitive to above
10 mBq/m$^3$ of radon in the water as shown in Fig.~\ref{rn_water}.
However, an effort has been made to attempt precise measurement of 
the background levels of those sensors.
The current background may include a few mBq/m$^3$ 
contamination from the radon sensor environment.
Therefore, we obtained only upper limits (1$\sigma$) for 
radon concentrations in water as listed in Table~\ref{rn_data}.
The current radon concentration in the SK tank water is less than
5.7 mBq/m$^3$. 
It is roughly 100 times lower than that in the water of Kamiokande-III.

\section{Test runs with radon-rich water}

In order to estimate the radon concentration in the SK water
independent of the radon sensor measurements, we used the Super-Kamiokande 
to take data with a known amount of radon in December 1997.

A total 1189 m$\ell$ of radon-rich water, equivalent to 13 Bq of radon,
was deposited in the central region of the SK detector 
via a 4 mm$\phi \times $ 22.0 m length rigid nylon tube on December 18, 1997.
The radon-rich water was made by the bubbling technique developed by 
our group\cite{nim_radon}.
The systematic error for this radon concentration is estimated to
be 11\% according to the systematic error for 
the accuracy of the liquid scintillation counter and 
the ionization chamber reported in Ref.~\cite{nim_radon}.
The water circulation was stopped from December 7, 1997
to January 17, 1998 in order to prevent the water flow
from stirring the supplied radon. 
We categorized the runs taken as listed in Table~\ref{category}.

To analyze the data of these test runs, both Super-Low-Energy (SLE) events 
and normal Low-Energy (LE) events are used.
LE events have been used for the solar neutrino analysis above 6.5 MeV,
which is described in Ref.~\cite{skflux}. 
The SLE events have energies between 5.5 MeV and 6.5 MeV.
SLE events have been taken since May 1997 using a software trigger.
When the hardware trigger threshold at 50\% efficiency was lowered 
from 5.7 MeV (for LE) to 4.6 MeV (for SLE), the raw event rate increased
from 10 Hz to 120 Hz. Most of the extra events occur near the detector wall.
Therefore, a fast vertex reconstruction and a preliminary fiducial volume cut
are applied to all the events in real-time on a workstation.
After this fiducial volume cut, the event rate becomes about 20 Hz.
Then all the events are stored to a magnetic tape library.
For this analysis, the same event selections as the SK solar neutrino
analysis\cite{skflux} are applied to the SLE and LE events.
Some additional noise cuts are also applied using the quality of 
the reconstructed vertex.

\begin{figure}[b]
\center
\includegraphics[width=12cm,clip]{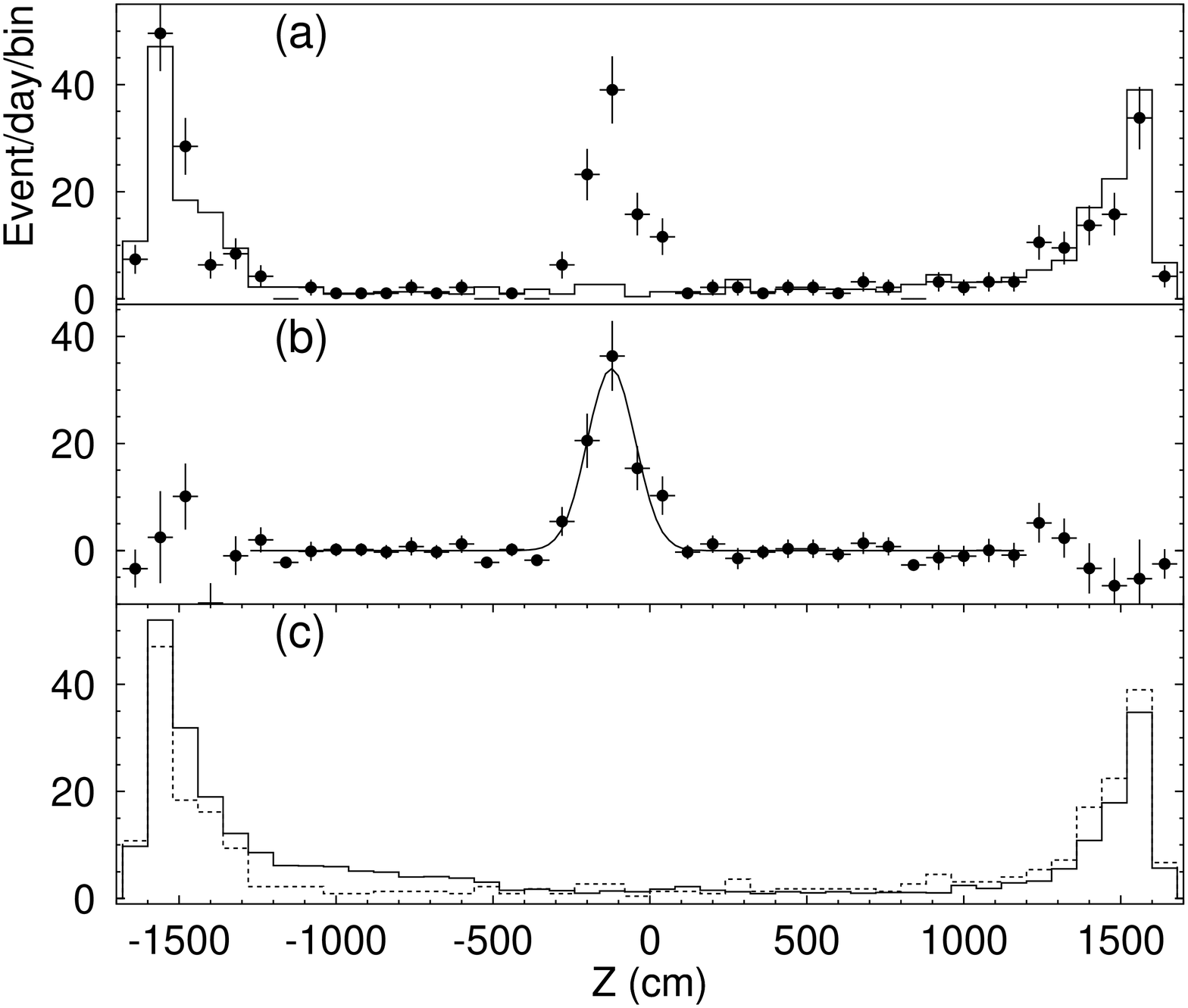}
\caption{
}
\label{rn_run}
\end{figure}

Figure~\ref{rn_run}(a)(c) shows reconstructed vertex position distributions of 
these SLE and LE events along the cylindrical axis (Z-axis). 
A fiducial volume in this plot is chosen as a cylindrical volume of 700 cm 
in radius and 3220 cm in height at the central region of the SK tank. 
The high event rates at both ends of the Z-axis are 
due to incoming gamma-ray events from the detector wall.
For the solar neutrino analysis, these gamma-ray backgrounds are 
eliminated.
The events in Fig.~\ref{rn_run} have energies between 5.0 MeV and 6.5 MeV.
The Rn-run, plotted with solid circles in Fig~\ref{rn_run}(a),
shows a peak in the Z-vertex distribution around the central region 
of the SK detector. 
This peak corresponds to events induced by the introduction of 
the radon-rich water.
The histogram in Fig~\ref{rn_run}(a) shows the vertex position distribution for
the BG-run. The vertical axes are normalized by run time.
Figure~\ref{rn_run}(b) was obtained by subtracting 
the BG-run from the Rn-run.
Applying a Gaussian fit to the distribution and calculating the excess 
number of events, we obtain
the detection efficiency of identifying these radon events as
$ 8.5 \times 10^{-5}$ around the central region of the detector.

Figure~\ref{rn_run}(c) is the vertex position distribution for the Normal-run
(solid line) and BG-run (dashed line).
Using the event rate in the Normal-run and the detection efficiency, 
we measure the radon concentration in the SK detector:
\begin{center}
\begin{tabular}{c l}
 $<$ 1.4 mBq/m$^3$  & (Z= 0 $\sim$ 1000 cm) , \\
   5 mBq/m$^3$  & (excess at Z = --1100 cm), and \\
   3 mBq/m$^3$  & (excess at Z = --600 cm).
\end{tabular}
\end{center}
In the Z $\geq 0$ cm region, the observed event rate corresponds 
to 1.4 mBq/m$^3$ of radon, but all the events are not necessarily 
due to radon. 
Therefore, an upper limit of radon concentration was obtained for this region,
and is consistent with the limit measured by the radon monitoring system.

In the bottom region of the SK detector, a higher event rate was observed
below 6.5 MeV.
We believe this excess to be caused by radon emanation from 
the detector wall, i.e. 20-inch PMT glass,
black plastic sheets, and/or stainless steel frames.
The water inlets to the SK tank are arranged at the bottom,
and the upward water-flow stirs radon from the materials 
at the detector wall into the bottom region of the SK detector. 
Temperature and particle count data indicate the
existence of flow cells with a sharp boundary at Z=--6 m.
This causes the event excess around the bottom region of the SK detector. 
From those test runs, the radon concentration 
in the bottom region is estimated as $3 \sim 5$ mBq/m$^3$ 
The radon emanation rate from 20-inch PMTs has been measured\cite{pmt_radon},
and is consistent with the order of magnitude of the observed 
radon concentration.

\section{Conclusions}

The radon concentrations at Super-Kamiokande have been measured
by using high sensitivity radon sensors.
The measured radon concentrations in the mine air and in the purified air
are $40 \sim 1200$ Bq/m$^3$ and $2 \sim 3$ mBq/m$^3$, respectively.
In the purified water, the radon concentration 
measured by the high sensitivity radon sensors is less 
than 5.7 mBq/m$^3$ in the SK tank.

Radon-rich water was supplied to the central region of the
SK detector in order to use SK low-energy events to
estimate the radon concentration in the SK water.
The current low-energy event rate in the top half of the SK detector
implies a radon concentration less than 1.4 mBq/m$^3$.
It is consistent with the value measured by the high sensitivity 
radon sensors for water.
The radon concentration in the bottom region is estimated as 
$3 \sim 5$ mBq/m$^3$.
We believe these radon is emanated from the detector materials, 
and then is stirred up by the upward water flow.

Relative to Kamiokande-III, we have achieved 100 times lower 
radon concentration in the SK tank water.
It allows us to lower the energy threshold for the solar 
neutrino analysis from 7.0 MeV (at Kamiokande-III) to 5.5 MeV (at current SK).

\section{Acknowledgments}

The authors acknowledge the cooperation of the Kamioka Mining and 
Smelting Company.
The Super-Kamiokande has been built and operated from funding 
by the Japanese Ministry of Education, Science, Sports and Culture,
and the United States Department of Energy.
This work was supported in part by Grant-in-Aid for Scientific 
Research (B) of the Japanese Ministry of Education, Science and Culture.

\newpage

\begin{center}
    \Large\bf Figure captions
\end{center}

Figure 1: \\
\begin{quote}
 A schematic view of the water purification system and
 locations of radon sensors.
\end{quote}

Figure 2: \\
\begin{quote}
 Time variation of radon concentration in air in various locations.
 A new radon trap device was introduced in March 1998.
 The increase in radon in the purification room air for the
 summer months reflects a temperature-driven reversal of air
 flow in the mine.
 The bump in radon in SK-tank air in June, 1998 was due to a malfuction of 
 the air purification system.
\end{quote}

Figure 3: \\
\begin{quote}
Time variation of radon concentrations in the SK-tank water and 
low-energy event rate from April 1996 to July 1996.
They show a strong correlation to each other.
\end{quote}

Figure 4: \\
\begin{quote}
Vertex position distributions for low-energy ($5 \sim 6.5$ MeV) events
in the SK detector during the test run period with radon-rich water.
(a): Rn-run (solid circles) and BG-run (histograms).
(b): vertex position distribution of Rn-run after subtracting the vertex
position distribution in BG-run.
(c): Normal-run (solid line) and BG-run (dashed line).
\end{quote}

\newpage
\begin{center}
    \Large\bf Tables
\end{center}
\vspace{2cm}

\begin{table}[h]
\begin{center}
\begin{tabular}{l c}
\noalign{\hrule height0.8pt}
Measurement point	& Radon concentration \\
\hline
Input water & $<$ 3.2 mBq/m$^3$ \\
Return water & $<$ 5.0 mBq/m$^3$ \\
SK tank water & $<$ 5.7 mBq/m$^3$ \\
\noalign{\hrule height0.8pt}
\end{tabular}
\vspace{5mm}

\caption{Averaged radon concentration in water in January, 1998.
The obtained values are 1$\sigma$ upper limits. (see text)}
\label{rn_data}
\end{center}
\end{table}

\vspace{1cm}

\begin{table}[h]
\begin{center}
\begin{tabular}{l c c c}
\noalign{\hrule height0.8pt}
Category & Run number & Date & Live time [day] \\
\hline
Normal-run   & R5200--5333 & Nov. 6 -- Dec. 8  & 15.4 \\
BG-run      & R5360--5376 & Dec. 15 -- Dec. 18 &  2.2 \\
Rn-run      & R5377--5380 & Dec. 18 -- Dec. 19 &  0.95 \\
\noalign{\hrule height0.8pt}
\end{tabular}
\vspace{5mm}

\caption{Test runs for the radon measurement.
Normal-run is usual data taking when the water purification 
system is runing.
BG-run is data taking with the water system stopped.
Rn-run is data taking after supplying the radon-rich water.
}
\label{category}
\end{center}
\end{table}

\newpage
\begin{center}
    \Large\bf Figures
\end{center}

\end{document}